\documentclass[a4paper,10pt,twoside]{cpc-hepnp}

\usepackage{multicol}
\usepackage{graphicx}
\usepackage{booktabs}
\usepackage{amssymb,bm,mathrsfs,bbm,amscd}
\usepackage[tbtags]{amsmath}
\usepackage{lastpage}
\usepackage{CJK}

\begin{document}

\title{Work of a multi-cathode counter in a single electron counting mode}

\author{%
\quad A.V.Kopylov $^{1}$\email{beril@inr.ru}%
\quad I.V.Orekhov \quad V.V.Petukhov }

\maketitle

\address{%
Institute for Nuclear Research of the Russian Academy of Sciences, \\
Moscow, 60-th October Anniversary prospect, 7A,  117312,  Russia}

\begin{abstract}
We describe a work of a multi-cathode counter of the developed
design in a single electron counting mode with a cathode made of
aluminum alloy. The results of the calibration of the counter are
presented. The coefficient of gas amplification was found from the
calibration spectra. The electric fields and operation of this
detector in two configurations are described and the original idea
to find the effect from electrons emitted from the surface of a
cathode by difference of the rates measured in two volume
configurations is expounded. Furthermore, the advantage of using a
multi-cathode counter for measurement of the intensity of single
electron emission from a metal is explained.
\end{abstract}

\begin{keyword}
proportional counter, single electron counting, low threshold
detector
\end{keyword}

\begin{pacs}
29.40.Cs
\end{pacs}

\begin{multicols}{2}

\section{Introduction}

At present time in experiments like the search for dark matter,
observation of coherent elastic neutrino-nucleus scattering et al
the low threshold detectors are used with high efficiency in the
detection of low energy ionizing radiation. One of the background
sources in these detectors are single electrons emitted from a
metallic surface. In a number of experiments \cite{1}--\cite{4} the
emission rates of single electrons from a photocathode of PMT have
been measured at different temperatures from a cryogenic one till a
room temperature. Contrary to the expectations the emission rate is
not decreased by lowering the temperature but instead, the
measurements have shown that it has been increased substantially.
This unexpected phenomenon still did not find a satisfactory
explanation. In the measurements with PMT the multi-alkaline
cathodes have been used with a low work function of less than 1 eV.
Because in many detectors the metals with relatively high (of about
4 $\div$ 5 eV) work function are used, it looks expediently to
measure the emission rates for different metals.  Since the emission
rates expected for these metals are lower than for photocathode of
PMT one needs a detector with higher sensitivity to single electrons
and with low dark rate. One way to get it is to use a gaseous
detector with a relatively large metallic surface. To register
single electrons detector should have high gas amplification. But
these demands: high surface, low dark rate, high gas amplification
look somewhat contradictory.  To satisfy all these demands a special
multi-cathode counter has been developed by us. A short description
and preliminary results of measurements by using this counter with a
copper cathode have been published in \cite{5}, \cite{6}. Here we
describe in detail the work of the counter of the developed design
with a cathode made of an aluminum alloy and focusing rings to
diminish effect from the ends of the counter.

\section{Design and method}

The picture of the counter and its schematic drawing are presented
at Fig. 1. The counter is placed in hermetic stainless steel
housing. All electrical connections are made by means of the
feed-through at one end of the counter.

\begin{center}
\includegraphics[width=6cm]{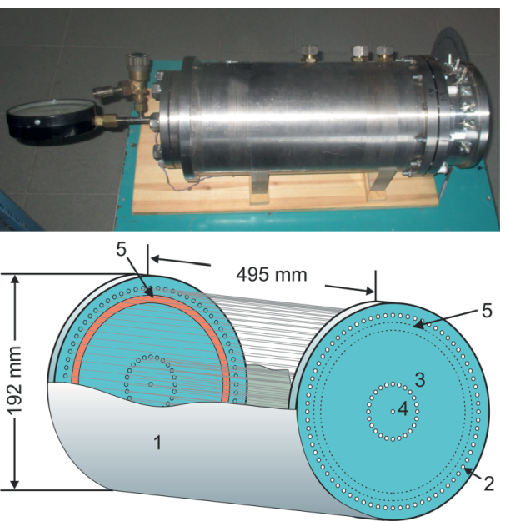}
\figcaption{\label{fig1} Top -- the picture of the counter. Bottom
-- schematic drawing of a multi-cathode counter.  1 -- an aluminum
cathode, 2 -- second cathode, 3 -- third cathode, 4 -- anode, 5 --
focusing rings.}
\end{center}

Counter has high sensitivity for the detection of single electrons
emitted from a cathode due to relatively large ($\approx$ 3000
$cm^2$) surface of the external cathode. At a distance 5 mm from an
external cathode a second one is mounted by an array of the nichrome
wires of a diameter 50 $\mu m$ tighten with a pitch 4.5 mm.
Electrons emitted from the surface of an external cathode are
drifting towards a central counter with a third cathode of 40 mm in
diameter assembled by the nichrome wires tighten with a pitch 6 mm.
Anode of a central counter is made of gold-plated W-Re wire of
diameter 25 $\mu m$. This geometry enables to get a relatively high
coefficient of gas amplification of about $10^5$. As a working gas a
mixture of $Ar + 10\%CH_4$ has been used at a pressure 0.1 MPa. At
the ends of the counter the focusing rings under potential of a
second cathode have been used to shift the trajectories of drifting
electrons from the ends of the counter towards its center. This
prevents the absorption of the electrons emitted from a cathode by
the ends of the counter. To shield a detector from an external
$\gamma $--radiation the counter has been placed in an iron cabinet
of 30 cm thick. The measurements have been performed at sea level at
a ground floor in Troitsk, Moscow region. High voltage from three HV
sources has been applied to cathodes 1, 2 and 3: $U_1$ = - 2500 V to
a cathode 1, $U_2$ = - 2480 V to a cathode 2 in a first
configuration and $U_2$ = - 2525 V to a cathode 2 in a second
configuration, $U_3$ = - 1750 V to a cathode 3. Figure 2 shows the
potentials, figure 3 -- field strength at the edge of the counter.
The calculations have been done by ANSYS Maxwell. One can see from
comparison of these figures the effect of a barrier from the second
cathode in configuration 2.

\begin{center}
\includegraphics[width=6cm]{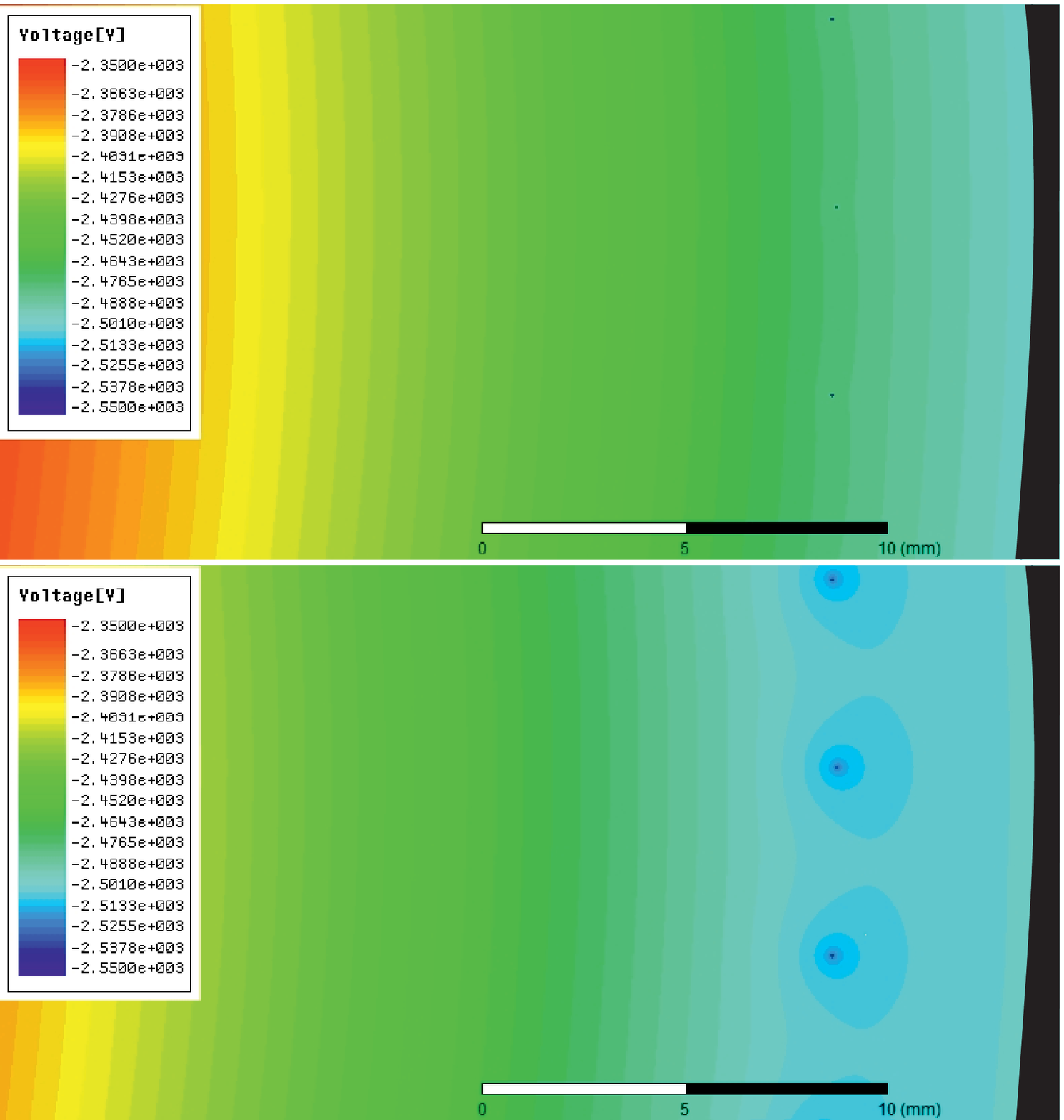}
\figcaption{\label{fig2} The potentials in configurations 1 (top)
and 2 (bottom)}
\end{center}

\begin{center}
\includegraphics[width=6cm]{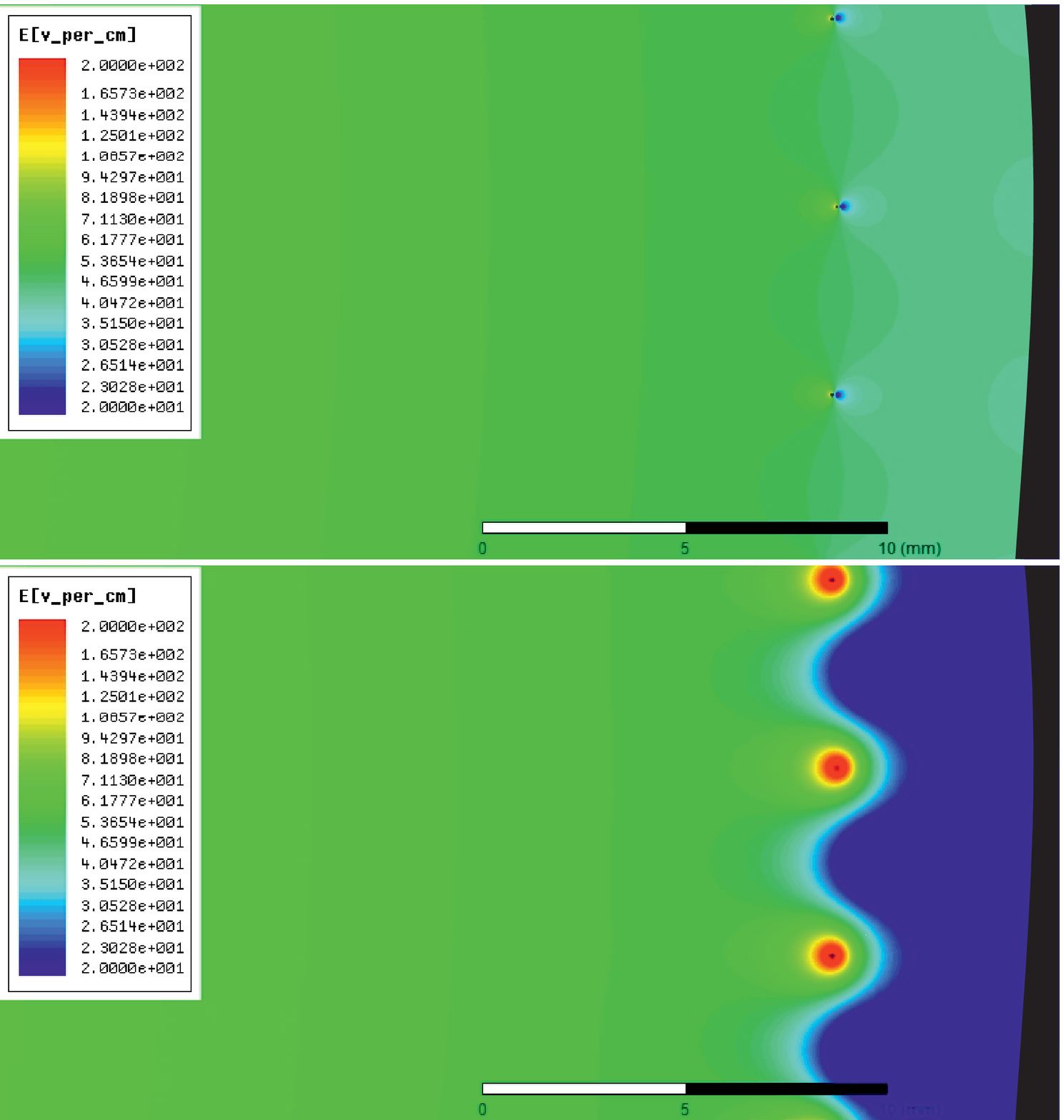}
\figcaption{\label{fig3} The field strength in configuration 1 (top)
and 2 (bottom)}
\end{center}

A signal from a central wire is fed to an input of a charge
sensitive preamplifier with a sensitivity of about 0.36 V/pC. The
output of preamplifier is connected to the input of 8 bit digitizer
NI5152. The shapes of the pulses have been digitized in the interval
$\pm$ 50 mV with a frequency 10 MHz and a digitizing interval 400
$\mu V$. The measurements have been collected round the clock; the
treatment of the obtained data has been performed off-line. The data
have been collected in two configurations of the counter: in a first
configuration the potential $U_2$ = - 2480 V applied to a second
cathode enables electrons emitted from an external cathode freely
drift to a central counter, in a second configuration a retarding
potential $U_2$ = - 2525 V is applied to a second cathode preventing
electrons to move freely towards a central counter. As a measure of
the effect from single electrons emitted from a cathode the
difference $\Delta R = R_1 - R_2$ of the count rates measured in a
first and second configurations was taken. Here it was assumed that
the rate $R_1$ is composed of the single electrons emitted from an
external cathode plus electrons emitted from wires and from the ends
of the counter while the rate $R_2$ is composed only of electrons
emitted from wires and from the ends of the counter. The novel idea
in using the difference $\Delta R = R_1 - R_2$ for the evaluation of
the effect from the surface of a cathode was to subtract two volume
effects of the counter obtained in configurations 1 and 2 with close
geometries which enables to subtract the background from wires of
the cathodes and from the ends of the counter. In other words,
instead of using ``fiducial volume'' as it has been typically done
in many of experiments, here we used ``fiducial surface'' to measure
the emission rate of single electrons emitted from a cathode of the
counter.

\section{The calibration of the counter}

The counter was calibrated by UV Mercury vapor lamp through a quartz
window in the side of the counter. Single electrons are emitted by
the absorption of UV light on the surface of aluminum cathode.
Figure 4 shows the shapes of the single electron pulses on the
output of CSP.

\begin{center}
\includegraphics[width=8cm]{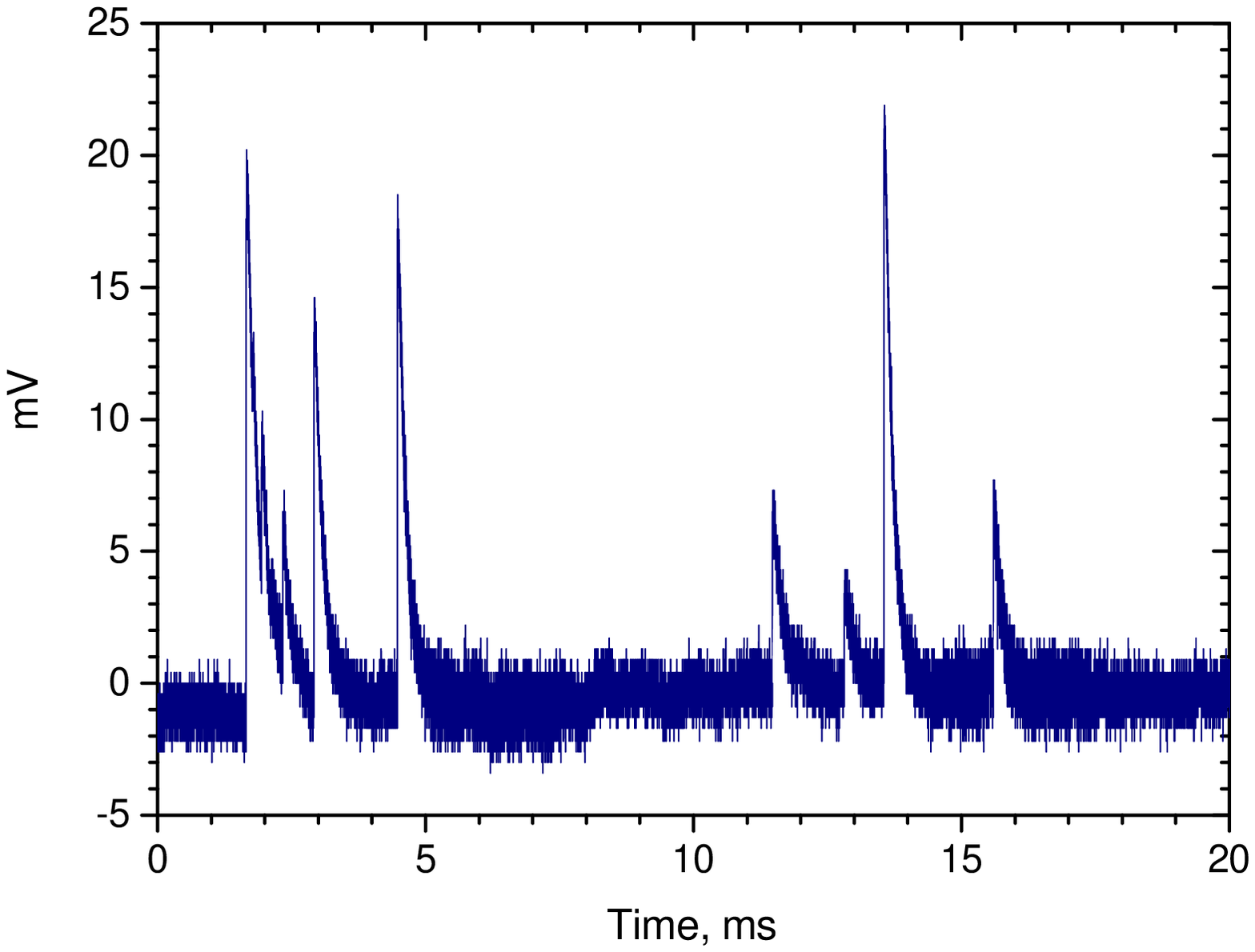}
\figcaption{\label{fig4} The pulses from single electrons on the
output of charge sensitive preamplifier generated during calibration
by UV Mercury vapor lamp.}
\end{center}

One can see that single electron pulses are well observed above
electronic noise which was measured to be $\approx 700 \mu V$ (RMS).
The interval of the amplitudes of the pulses used in measurements in
both configurations was taken from 3 to 30 mV. Fig.5 shows the
calibration spectra obtained in configuration 1 and 2. The dark rate
measured by closed window has been subtracted from the effect
measured by opened window.

\begin{center}
\includegraphics[width=8cm]{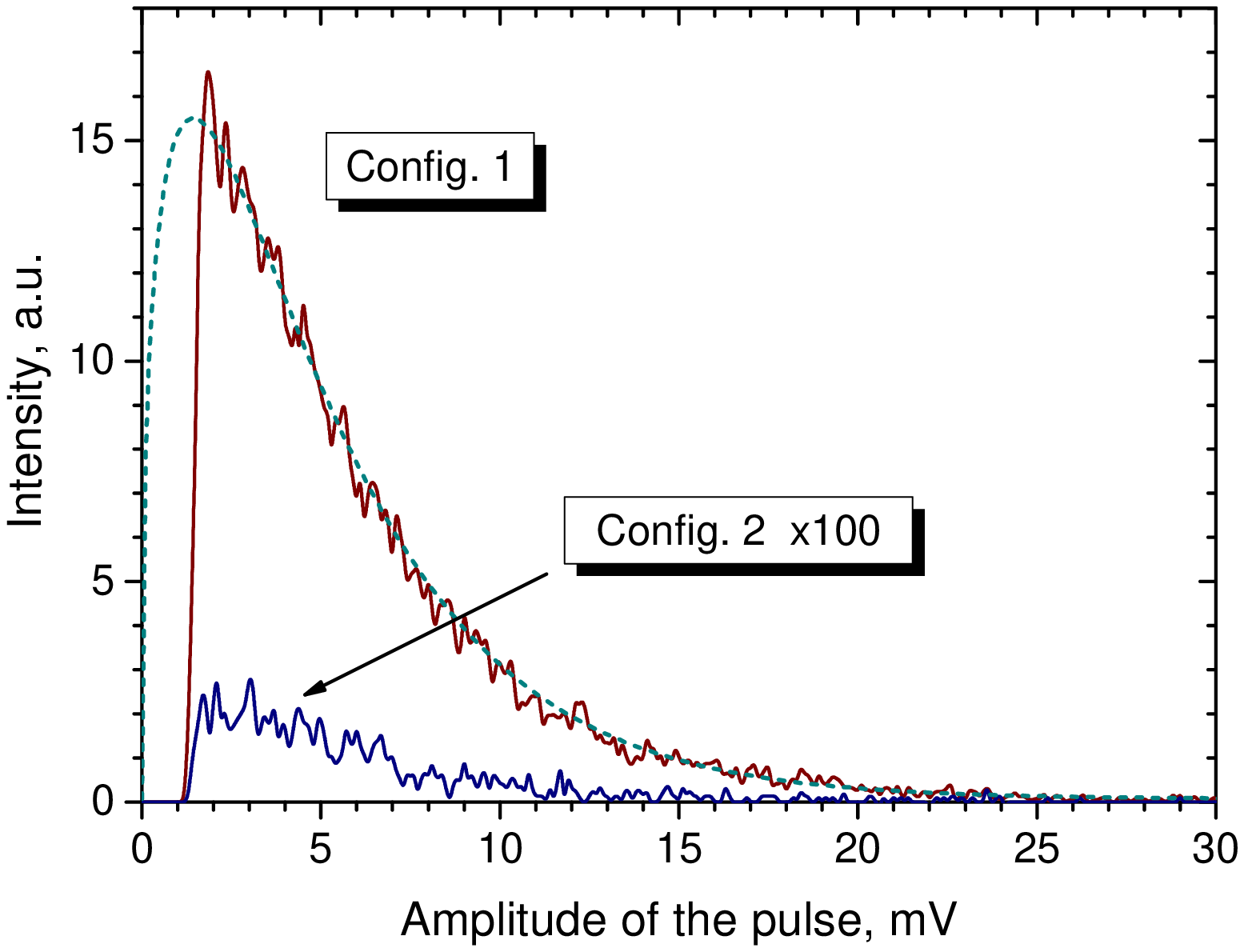}
\figcaption{\label{fig5} The calibration spectra of single electron
events obtained in configurations 1 and 2. Dashed line -- Polya
distribution.}
\end{center}

One can see from the comparison of these two spectra, obtained by
the same intensity of UV Mercury vapor lamp that in configuration 2
electrons emitted by UV light from the surface of the cathode of the
counter do not pass a barrier of the cathode 2 and hence, are not
registered by a central counter. This proves that the observed
effect is most probably from single electrons emitted by UV light
from the surface of the external cathode. As it is well known the
spectrum of single electron pulses in gaseous detectors  is
described by Polya distribution:

\begin{eqnarray}
\label{eq1}
P\left(A\right)=C{\left(\frac{A}{\bar{A}}\right)}^{\theta
}e^{-\left(\left(1+\theta \right)\frac{A}{\bar{A}}\right)}
\end{eqnarray}

Here $A$ is the amplitude of the pulse, $\bar{A}$ is the mean
amplitude and $\theta $ -- is a parameter which depends on the
counting characteristics. For this counter $\theta $ was found to be
equal $\approx 0.5$. The rate was taken as an integral by the
interval of the amplitudes of the pulses from 3 to 30 mV. In this
interval Polya distribution is well approximated by exponent. The
efficiency calculated from this distribution and interval from 3 to
30 mV was found to be ($60 \pm 2$)\%. The average amplitude of the
pulse found from Polya distribution is 5 mV. The sensitivity of CSP
was 0.36 V/pC. One can easily find from this that gas amplification
of our detector is about $10^5$. The pulses were selected also by
prehistory of the event as it was described in \cite{6}. The events
with the strong ($>5 mV$) deviation of a base line from zero were
discarded. With all discriminations the live time of the counting
procedure was found to be ($70 \pm 5)$\%. Fig.6 shows the count
rates measured by calibration for different potentials applied to a
cathode 2.

\begin{center}
\includegraphics[width=8cm]{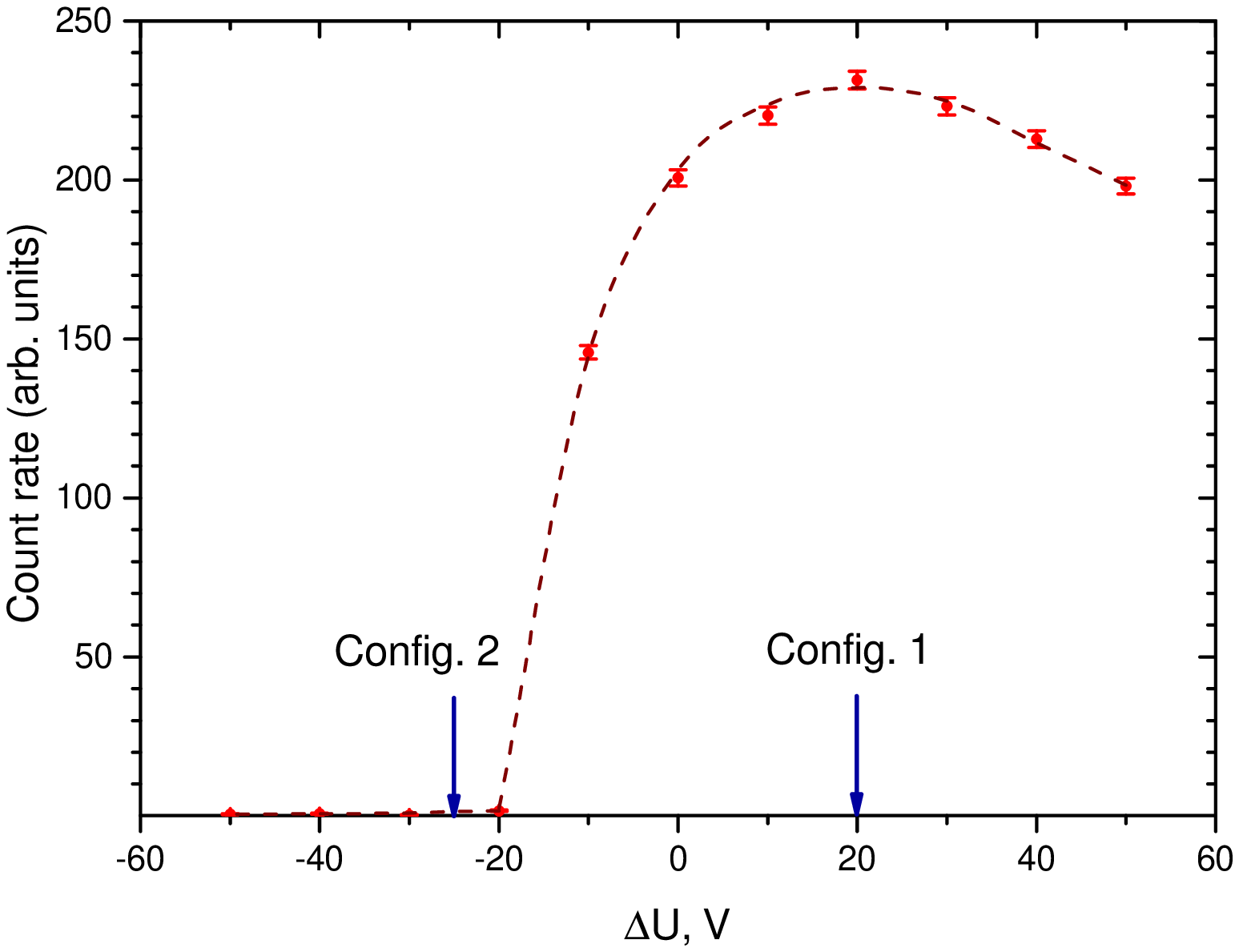}
\figcaption{\label{fig6} Count rate by calibration of the counter as
function  $\Delta U = U_2 - U_1$. By arrows are shown the optimal
potential difference for configurations 1 and 2.}
\end{center}

One can see, that at $\Delta U$ =  20 V the count rate is maximal.
At higher $\Delta U$ the rate falls down because some portion of
electrons emitted from the surface of external cathode gets caught
by a cathode 2. At lower $\Delta U$ some portion of emitted
electrons is scattered back to the external cathode. For
configurations 1 and 2 $\Delta U$ were taken as +20 V and -25 V
correspondingly as the optimal ones.

\section{Summary}

This paper is focused on the description of a design and work of a
multi-cathode counter in single electron counting mode. The use of
multiple cathodes enables to subtract efficiently the background
from the ends of the counter. Another advantage of using this design
is that it is very efficient technique of reducing the background
from single electrons emitted from external cathode by means of a
second cathode made of wires tighten close to the external cathode.
The calibration of the counter and measurement of the count rates at
different potentials applied to a second cathode enables to find the
optimal potentials applied to a second cathode in configurations 1
and 2. As a result this detector can be efficiently used to measure
the count rate of single electrons emitted from metal. Due to large
surface of the external cathode the detector has high sensitivity to
single electrons emitted from a cathode which makes possible to
measure the rate of emission of single electrons from metals with
relatively high work function like nickel or platinum. Because these
materials can be very promising for future detectors it is also in
our plans to assemble counters with a cathode made of these metals.

\acknowledgments{The authors express deep gratitude to E.P.Petrov
for substantial contribution to this work, particularly in
fabrication of the counter.}

\end{multicols}

\clearpage


\vspace{3mm}

\begin{thebibliography}{90}

\bibitem{1}James P.Rodman and Harlan J.Smith Applied Optics, 1963, {\bf 2}:
181

\bibitem{2}J.A.Nikkel,W.H.Lippincott and D.N.McKinsey JINST, 2007, {\bf 2}:
P11004

\bibitem{3}Y.O.Meyer EPL, 2010, {\bf 89}:
58001

\bibitem{4}Clara Cuesta (DUNE collaboration) arXiv: 1711.08307

\bibitem{5}A.V.Kopylov, I.V.Orekhov, V.V.Petukhov, Tech.Phys.Lett., 2016, {\bf 42}:
102

\bibitem{6}A.V.Kopylov, I.V.Orekhov, V.V.Petukhov, Advances In High Energy Physics, 2016:
2058372

\end{thebibliography}
\end{document}